\documentclass[preprint,aps,12pt,preprintnumbers,eqsecnum,nofootinbib]{revtex4}
\usepackage{graphicx}
\usepackage{subfigure}

\usepackage{color}
\usepackage{amssymb,amsmath,amsfonts}
\usepackage{epstopdf}

\def\rhotil{{\tilde{\rho}}}
\def\thetil{{\tilde{\theta}}}
\newcommand{\al}[1]{\begin{align}#1\end{align}}

\newcommand{\beq}{\begin{equation}}
\newcommand{\enq}{\end{equation}}

\begin{document}
%
\title{\vspace*{0.5in} 
Flavored axion-monodromy inflation
\vskip 0.1in}
\author{Christopher D. Carone}\email[]{cdcaro@wm.edu}
\author{Raymundo Ramos}\email[]{raramos@email.wm.edu}
\author{Zhen Wang}\email[]{zwang01@email.wm.edu}

\affiliation{High Energy Theory Group, Department of Physics,
College of William and Mary, Williamsburg, VA 23187-8795}
%
%
\date{September 26, 2015}

\begin{abstract}
The hierarchy of fermion masses in the standard model may arise via the breaking of discrete gauge
symmetries.  The renormalizable interactions of the flavor-symmetry-breaking potential can have accidental
global symmetries that are spontaneously broken, leading to pseudo-goldstone bosons that may drive inflation.
We consider two-field, axion-monodromy inflation models in which the inflaton is identified with a linear
combination of pseudo-goldstone bosons of the flavor sector.  We show that the resulting models
are nontrivially constrained by current cosmological data as well as the requirements of viable flavor model 
building.
\end{abstract}
\maketitle

\section{Introduction} \label{sec:intro}

The prevailing approach to solving the horizon and flatness problems of conventional Big Bang cosmology
is inflation, a period in which the universe underwent exponential expansion due to the effects of the
nearly constant energy density provided by a scalar field~\cite{inflate}.  Models of inflation are often studied in terms
of the properties of the inflaton potential, with somewhat less focus on other roles the inflaton might play in
extensions of the standard model.  If the inflaton has no purpose other than to provide the source of the energy
density that drives inflation, then model building becomes isomorphic to studying ways of generating different
functional forms for the inflaton potential.  These possibilities, now cataloged (see for example~\cite{Lyth:2007qh}), differ 
in their detailed predictions for the spectrum of fluctuations in the microwave background that are observed in experiments like 
Planck~\cite{Ade:2015lrj} and BICEP2~\cite{Ade:2014xna}.    

In this paper, we consider a scenario in which the inflaton is an integral component of an extension of the standard 
model that aims to address one of its substantial mysteries: the hierarchy of elementary fermion masses.  
Models of flavor based on horizontal discrete symmetries postulate that these symmetries are broken via a set of 
fields, called flavons, that couple to standard model fermions through higher-dimension operators.  Discrete flavor
symmetries can often lead to accidental continuous global symmetries among the renormalizable terms of the flavon potential.  
In the present work, we consider the possibility that the inflaton may be identified as a linear combination of the approximate 
goldstone bosons that arise when these accidental symmetries are spontaneously broken.  We will ultimately be interested in 
two-field models of inflation, for reasons described below, which distinguishes the present work from the relatively sparse 
literature that explores the use of flavon fields for a similar purpose~\cite{Antusch:2008gw}.

Consider the simplest possibility, a $\mathbf{Z}_N$ flavor symmetry under which a single flavon field $\Phi$ transforms 
as $\Phi \rightarrow \omega \Phi$, where $\omega = \exp(2 \pi i /N)$.  If the fermions of the standard model are charged under the discrete 
group, then a tree-level Yukawa coupling that would otherwise be forbidden can arise via a higher-dimension operator.  For example,
for a down-type quark, one might have
\begin{equation}
\frac{1}{M_F^p} \bar{Q}_L H \phi^p D_R + \mbox{h.c.} \,\,\, ,
\end{equation}
where $H$ is the standard model Higgs doublet, $M_F$ is the flavor scale, and $p$ is an integer.   The Yukawa coupling is 
associated with the ratio $(\langle \phi \rangle/M_F)^p$ which can be much less than one; operators with different values of $p$ 
can easily provide a hierarchical pattern of entries in the associated Yukawa matrix.  If less than the Planck scale, the scale $M_F$ is 
identified with that of new heavy states that account for the origin of the higher-dimension operators.  However, a simpler 
assumption, that we adopt henceforth, is that the scale $M_F$ is the reduced Planck scale $M_*$; the desired operators appear 
as part of the most general set that are allowed by the local symmetries of the theory, as one expects based on our current understanding
of quantum gravity~\cite{Kallosh:1995hi}.   An immediate implication of our assumptions is that the  vacuum expectation value (vev) $\langle \Phi \rangle < M_*$, which 
will provide an important constraint in our attempt to identify the inflaton with a part of the field $\Phi$.

To obtain an inflaton potential that is sufficiently flat, we require that the goldstone boson degree-of-freedom receives no contributions to
its potential from renormalizable terms involving $\Phi$.  Let us therefore assume that $N \geq 5$.  The renormalizable terms in the potential
are simply
\begin{equation}
V(\Phi) = -m_\Phi^2 \Phi^\dagger \Phi + \frac{\lambda_\Phi}{2} (\Phi^\dagger \Phi)^2 \,\,\, .
\label{eq:exvphi}
\end{equation}
Terms such as $(\Phi^4 + \mbox{h.c.})$ are forbidden by the $\mathbf{Z}_N$ symmetry.  Using the nonlinear decomposition
\begin{equation}
\Phi = \frac{\phi + f}{\sqrt{2}} \exp(i \theta/f)  \,\,\,,
\end{equation}
where $f/\sqrt{2} \equiv \langle \Phi \rangle$, one sees immediately that $V(\Phi)$ is independent of $\theta$, {\em i.e.}, the potential 
$V(\theta)$ is exactly flat.     The potential in Eq.~(\ref{eq:exvphi}) has an accidental global U(1) symmetry and the field $\theta$ is
the goldstone boson that results from its spontaneous breaking.  Global symmetries are not respected by quantum gravitational corrections, so
it is no surprise that there are Planck suppressed corrections,
\begin{equation}
{\cal L} \supset \frac{c_0}{2} \frac{1}{M_*^{N-4}}\, \Phi^N + \mbox{ h.c. } \,\,\, ,
\label{eq:exop}
\end{equation}
that generate a potential for $\theta$, where $c_0$ is an unknown order-one coefficient.  Planck-suppressed operators that directly break the 
discrete flavor symmetry are not present since we assume in this example (and will require in all our models henceforth) that we work with discrete gauge 
symmetries, which satisfy appropriate anomaly cancellation conditions and are immune to quantum gravitational corrections.  For the reader who is unfamiliar with discrete gauge symmetries, we review the basic issues relevant to our model building in the appendix.  

The operator in Eq.~(\ref{eq:exop}) leads to the $\theta$ potential
\begin{equation}
V(\theta) = c_0\, M_*^4 \left(\frac{\langle \Phi \rangle}{M_*} \right)^N \left[1-\cos\left( N \theta/f \right) \right] \,\,\, ,
\end{equation}
where we have added a constant so that $V(0)=0$.  This is nothing more than the potential of ``Natural Inflation"
scenarios~\cite{NI}.  However, this potential is not adequate for our purposes.  It is well known that if one requires that Natural Inflation provides $\sim 50-60$ 
e-folds of inflation and predicts a spectral index $n_s$ within the range allowed by current Planck data, then $f$ must be well above the Planck 
scale~\cite{Kim:2004rp}. For our present application, this would imply that $\langle \phi \rangle/M_*$ is not a small flavor-symmetry-breaking parameter and we lose the
ability to predict standard model Yukawa couplings in a controlled approximation.

We therefore must consider other ways of generating potentials for the pseudo-goldstone inflaton that allow a sub-Planckian decay content $f$.  The options 
assuming a single field inflation model are limited.  For example, models of ``multi-natural" inflation~\cite{Czerny:2014wza}, in which one arranges for additional 
sinusoidal terms in the potential, can accommodate a sub-Planckian flavon vev, but tend to predict $n_s = 0.95$ in this limit~\cite{Czerny:2014wza}, at the very 
edge of the 95\% exclusion region following from Planck data.  A different class of models that can more easily provide cosmological predictions consistent with Planck data 
are two-field models of the axion monodromy type~\cite{McAllister:2008hb,Berg:2009tg,Carone:2014cta,mcdonald,Barenboim:2014vea,Li:2015taa,Erlich:2015xya}.  We will show that these can be adapted for the present 
purpose.   

The two pseudo-goldstone fields can have their origin if there are two flavon fields, $\Phi$ and $\chi$, that transform under the discrete group 
$\mathbf{Z}^\Phi_p \times \mathbf{Z}^\chi_r$.  We assume that each field transforms only under one of the $\mathbf{Z}_N$ factors,
\begin{equation}
\Phi \rightarrow \omega_\Phi \Phi \,\,\,\,\,\,\,\,\,\, \mbox{ and } \,\,\,\,\,\,\,\,\,\, \chi \rightarrow \omega_\chi \chi \,\,\, ,
\label{eq:rule}
\end{equation}
where $\omega_\Phi = \exp(2 \pi i /p)$  and  $\omega_\chi = \exp(2 \pi i /r)$, where $p$ and $r$ are integers.  For $p\geq 5$ and $r \geq 5$, the
renormalizable terms in the potential are
\begin{equation}
V(\Phi,\chi) = -m_\Phi^2 \Phi^\dagger \Phi + \frac{\lambda_\Phi}{2} (\Phi^\dagger \Phi)^2 
-m_\chi^2 \chi^\dagger \chi + \frac{\lambda_\chi}{2} (\chi^\dagger \chi)^2 + \lambda_p \Phi^\dagger \Phi \chi^\dagger \chi \,\,\, ,
\label{eq:tfpot}
\end{equation}
where $\lambda_p$ is a portal-type coupling.   There is no difficulty in choosing parameters such that each field develops a vev.
This potential has an accidental U(1)$\times$U(1) global symmetry that is spontaneously broken.  Extending our previous parameterization,
we write
\begin{equation}
\Phi = \frac{\phi_0 + f_\theta}{\sqrt{2}} \exp(i \theta/f_\theta) \,\,\,\,\,\,\,\,\,\, \mbox{ and } \chi = \frac{\chi_0 + f_\rho}{\sqrt{2}} \exp(i \rho/f_\rho) \,\,\, .
\end{equation}
Spontaneous symmetry breaking renders the fields $\phi_0$ and $\chi_0$ massive so that they are decoupled from the inflation dynamics.  
The potential for the goldstone bosons $V(\rho,\theta)$ that follows from Eq.~(\ref{eq:tfpot}) is exactly flat.

We will discuss later how to generate a potential for $\rho$ and $\theta$ of the following axion-monodromy form  
\begin{equation}
V(\rho,\theta) = \Lambda_1^4 \left[1+ \cos \left(\frac{\rho}{f_\rho}\right)\right] +\Lambda_2^4\left[1- \cos \left(\frac{ n \, \rho}{f_\rho}-\frac{\theta}{ f_\theta}\right)\right] \,\,\,,
\label{eq:ourpot}
\end{equation}  
where $n$ is an integer. The first few terms in the expansion of the first cosine factor have the same form as  $-m_r^2 r^2/ 2+ \lambda_r r^4/4!$,  the 
shift-symmetry-breaking potential $W(r)$ assumed in the Dante's Waterfall scenario discussed in Ref.~\cite{Carone:2014cta}.  In that work, $W(r)$ was assumed to be 
generated by non-perturbative effects associated with moduli stabilization in string theory, as for example in Ref.~\cite{Berg:2009tg}.  In this paper, we only consider field 
theoretic origins of the potential, where the emergence of the functional form given in Eq.~(\ref{eq:ourpot}) is readily obtained.  For the purposes of graphical display, if one 
plots the potential as if $\rho$ and $\theta$ were polar coordinates, one would find a ``hill" generated by the first cosine factor, circumscribed by a descending 
spiral ``trench" generated by the second.  Inflationary trajectories track the minimum of the trench.   As $\theta$ advances by $2 \pi f_\theta$ along the trench, the $\rho$ 
coordinate does not return to  the same value; this monodromy allows for large numbers of e-folds to be achieved within a bounded, sub-Planckian region of field space.   
We assume that the decay constant 
$f_\theta$ satisfies
\begin{equation}
\frac{f_\theta}{ \sqrt{2}} = \lambda \, M_* \approx 0.22 \,  M_* \,\,\, ,
\label{eq:fchoice}
\end{equation}
where $\lambda$ is a flavor-symmetry-breaking parameter of the same size as the Cabibbo angle.  This will allow 
us to identify the field $\Phi$ (and perhaps in some models both $\Phi$ and $\chi$) as flavons that can be used in flavor model building.   We will see that the discrete 
symmetry $\mathbf{Z}^\Phi_p \times \mathbf{Z}^\chi_r$ serves four purposes: ($i$) it assures that there are goldstone bosons that have no potential generated 
by renormalizable couplings, ($ii$) it will serve as a flavor symmetry to create a hierarchy of standard model fermion Yukawa couplings, ($iii$) it will lead to the 
correct pattern of couplings in a new gauge sector that provides for the desired form of the inflaton potential, Eq.~(\ref{eq:ourpot}), and ($iv$) it will keep quantum
gravitational corrections to the potential highly suppressed.

Our paper is organized as follows.  In the next section, we discuss the inflationary dynamics that follows from the potential given in Eq.~(\ref{eq:ourpot}).  We identify 
solutions in which inflation ends when single-field slow-roll conditions are violated and other solutions where the termination of inflation is analogous to a hybrid 
model~\cite{Linde:1993cn}.  In Sec.~\ref{sec:fmodel}, we consider model building issues, in particular, how the discrete symmetries of the theory play an important role in assuring 
that we obtain the proper potential, and how the same symmetries can be used to produce a plausible model of standard model fermion masses.  In the final section,
we summarize our conclusions.  A brief appendix is provided to review relevant facts about discrete gauge symmetries.

\section{Inflatonary Trajectories} \label{sec:traject}

In this section, we consider inflationary trajectories in the two-field potential given by Eq.~(\ref{eq:ourpot}), that are compatible with flavor model-building 
requirement Eq.~(\ref{eq:fchoice}).  We give two example solutions that differ qualitatively in how inflation ends.  

\subsection{Termination without a waterfall.} \label{n21-nowaterfall} 

For our first solution, we make the parameter choice $f_\rho = f_\theta \equiv f_1$ and also define $f_1/n \equiv f_2$.  We assume $f_1 \gg f_2$, which is equivalent 
to $n\gg 1$.   The potential Eq.~(\ref{eq:ourpot}) then takes the form
\begin{equation}
  V(\rho,\theta) = \Lambda_1^4 \left[1+ \cos \left(\frac{\rho}{f_1}\right)\right]+\Lambda_2^4\left[1- \cos \left(\frac{\rho}{f_2}-\frac{\theta}{f_1}\right)\right] \,\, .
\end{equation}
The second cosine term creates a series of trenches on the surface of the potential defined by the first cosine term.  If the field $\theta$ is plotted as a
polar coordinate, the trenches form spirals originating at $\rho=0$.  As in Ref.~\cite{Carone:2014cta}, it is convenient to work in the rotated field basis 
$\rho = c \, \rhotil+ s\, \thetil$ and $\theta = c\, \thetil-s\, \rhotil$ with 
\begin{equation}
c=\frac{f_1}{\sqrt{f_1^2+f_2^2}}  \,\,\,\,\,\,\,\,\,\, \mbox{ and } \,\,\,\,\,\,\,\,\,\, s=\frac{f_2}{\sqrt{f_1^2+f_2^2}} \,\,\, .
\end{equation}
This allows us to rewrite the potential as
\begin{equation}
  V(\rhotil,\thetil) = \Lambda_1^4 \left[1+ \cos \left(\frac{c \rhotil+ s \thetil}{f_1}\right)\right]+\Lambda_2^4\left[1- \cos \left(\frac{\rhotil}{f}\right)\right],
  \label{eq:vtp}
\end{equation}
where $f=f_1f_2/\sqrt{f_1^2+f_2^2}$.  The modulations in the potential due to the $\cos(\tilde{\rho}/f)$ term create the trench, whose location is given by
$\partial V/\partial \rhotil=0$, or
\begin{equation}
\sin\left(\frac{\rhotil}{f}\right)-s \, c\, \frac{\Lambda_1^4}{\Lambda_2^4}\sin\left(\frac{c \rhotil+ s \thetil}{f_1}\right)=0 \,\,\,.
\label{firstder1}
\end{equation}
The inflaton is the linear combination of the fields that slowly rolls along the trench; inflation terminates when the slow-roll conditions are violated.  For the solutions 
considered in this subsection, the stability condition $\partial^2 V/\partial \rhotil^2 > 0$ will hold throughout this trajectory.

To study inflationary observables, we first consider a good approximation to the single-field inflaton potential, which holds for our choice of parameters and can
be studied analytically, and then discuss an exact numerical approach that we use to confirm the validity of our results.   Let us define 
$\kappa \equiv s \,c\, (\Lambda_1^4/\Lambda_2^4)$ and consider parameter choices where $\kappa \ll 1$.  It follows from Eq.~(\ref{firstder1}) that to good approximation
\begin{equation}
\tilde{\rho}/f \approx 2 \pi j \,\,\, ,
\label{eq:pgood}
\end{equation}
where $j$ is an integer.  Given our assumption that $f_1 \gg f_2$, it follows from Eqs.~(\ref{eq:vtp})-(\ref{eq:pgood}) that $\partial^2 V(\rhotil,\thetil)/\partial \rhotil^2 > 0$, confirming that
the trench is stable.  Substituting Eq.~(\ref{eq:pgood}) into our original potential yields
\begin{equation}
V(\tilde{\theta}) = \Lambda_1^4 \left[1 + \cos\left(\delta +\tilde{\theta}/f_0\right) \right] \,\,\, ,
\label{eq:appot}
\end{equation}
where $\delta =2 \pi  s c j $ and $f_0 = f_1/s$.   Setting $j=0$ is equivalent to redefining the origin of field space, so we will ignore $\delta$ henceforth.
We note that the present approximation scheme differs from the one used in Ref.~\cite{Carone:2014cta}, in which one would expand the sinusoidal functions
in Eq.~(\ref{firstder1}) to linear order in their arguments, but is nonetheless accurate as we confirm numerically later. We note that $s \ll 1$ in the limit $n \gg1$, 
so that the derived quantity $f_0$ can be super-Planckian even when the decay constants $f_1$ and $f_2$ are not.

\begin{figure}[t]
  \begin{center}
    \includegraphics[width=0.5\textwidth]{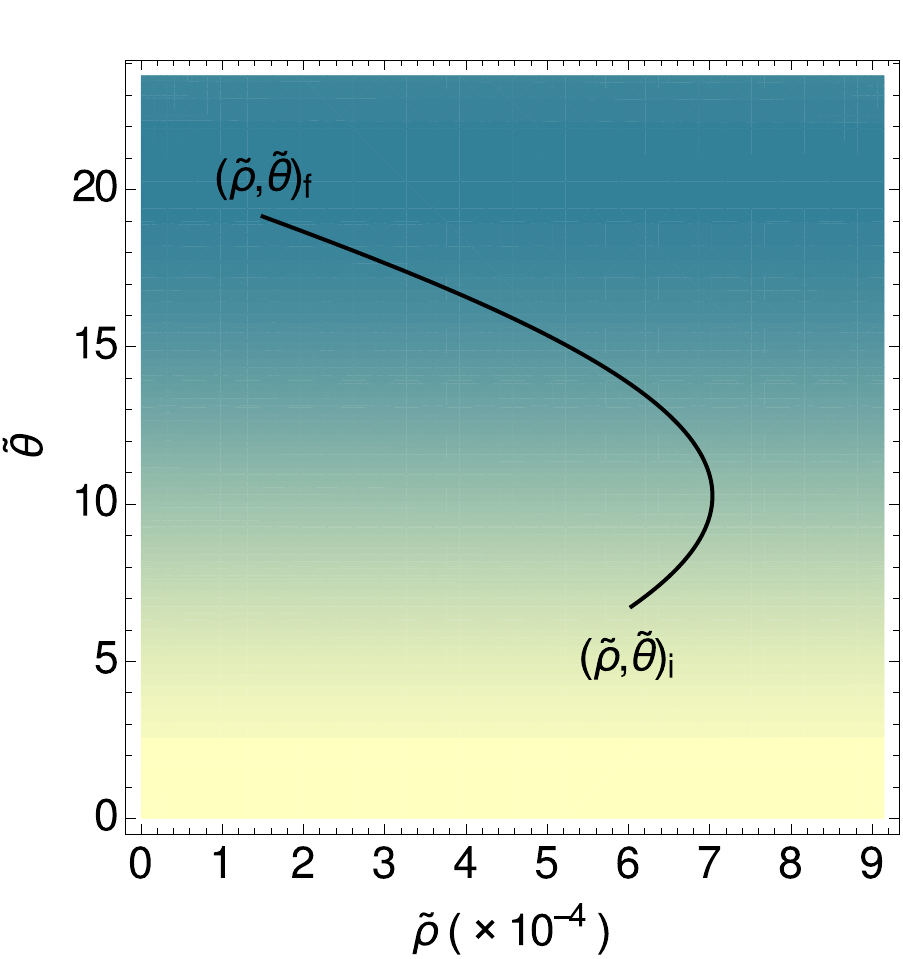}
    \caption{Path followed by the inflaton during $60$ e-folds of inflation corresponding to the solution of Sec.~\ref{n21-nowaterfall}, in units where $M_*=1$. 
    The background is a density plot where darker zones have lower values of the potential than lighter ones.}
        \label{fig:tilden21}
  \end{center}
\end{figure}

We compare the predictions of the model to the latest results from the Planck Collaboration \cite{Ade:2015lrj}. The slow roll parameters are defined by
\begin{equation}
\epsilon=\frac{1}{2} \left(\frac{V'}{V}\right)^2 \,\,\, , \,\,\,\,\,\,\,\,\,\, \eta=\frac{V''}{V} \,\,\,\,\, \mbox{ and } \,\,\,\,\, \gamma = \frac{V' V'''}{V^2} \,\,\, ,
\label{eq:srp}
\end{equation}
where the primes refer to derivatives with respect to the inflaton field and we work in units where the reduced Planck mass $M_* \equiv M_P/\sqrt{8\pi} =1$.  In the present case, 
these are given by
\begin{equation}
\epsilon = \frac{1}{2 f_0^2}\tan^2[\thetil/(2 f_0)] \,\,\,, 
\label{eq:ep1}
\end{equation}
\begin{equation}
\eta = -\frac{1}{f_0^2}\frac{\cos (\thetil/f_0)}{1+\cos (\thetil/f_0) } \,\,\, ,
\label{eq:et1}
\end{equation}
\begin{equation}
\gamma = -\frac{1}{f_0^4}\tan^2[\thetil/(2 f_0)] \,\,\, .
\label{eq:gm1}
\end{equation}
Inflation ends when $\epsilon(\tilde{\theta}_f)=1$.  The initial value of the inflaton, $\tilde{\theta}_i$ is determined
by the requirement that we achieve a desired number of e-folds of inflation, given in general by
\begin{equation}
  N=\int_{\thetil_i}^{\thetil_f} \!\! \frac{1}{\sqrt{2\epsilon}}\, d\thetil=2 f_0^2 \ln \!
   \left[\frac{\sin[\thetil_f/(2 f_0)]}{\sin[\thetil_i/(2 f_0)]}\right] \,\,\, .
\end{equation}
We set $N=60$ in the numerical results that follow.   We evaluate the slow-roll parameters  and the potential $V(\tilde{\theta})$ at $\tilde{\theta}_i$ in determining the spectral index $n_s=1-6\epsilon+2\eta$, the ratio of tensor-to-scalar amplitudes $r=16\epsilon$, the running of the spectral index $n_r=16\epsilon\eta - 24 \epsilon^2-2\gamma$ and the scalar amplitude $\Delta_R^2 = V/(24 \pi^2 \epsilon)$.  From Eqs.~(\ref{eq:ep1})-(\ref{eq:gm1}), it follows that
\al{
n_s &=1+ \frac{1}{f_0^2} \left(1-2 \sec ^2[\thetil_i/(2 f_0)]\right) \,\,\, ,\\
r & = \frac{8}{f_0^2}\tan^2[\thetil_i/(2 f_0)] \,\,\,\, ,\\
n_r &=  -\frac{2}{f_0^4} \tan^2[\thetil_i/(2 f_0)] \sec^2[\thetil_i/(2 f_0)] \,\,\, ,\\
\Delta_R^2 &= \frac{1}{12 \pi^2} \Lambda_1^4 f_0^2 \left(1+\cos[\thetil_i/f_0]\right)^3 \csc^2[\thetil_i/f_0] \,\,\, .
}
To illustrate a viable solution, consider the parameter choice (again, in units where $M_*=1$)
\al{
  f_1 &= 0.22 \sqrt{2}\,\,\, , \\
  f_2 &= f_1 / 21 \,\,\, ,\\  
  \Lambda_1 & =\Lambda_2 = 0.006 \,\,\, , 
}
which corresponds to $n=21$ and $\kappa \approx 1/21$.  We find that the initial and final fields for the inflaton trajectory are given by
\begin{equation}
   (\rhotil,\thetil)_i =(6.04\times 10^{-4},  6.74 ) \,\,\,\,\,\,\,\, \mbox{ and } \,\,\,\,\,\,\,\, (\rhotil,\thetil)_f =(1.50 \times 10^{-4}, 19.14) \,\,\, ,
\end{equation}   
respectively.  Using this value for $\tilde{\theta}_i$, we find the following set of cosmological parameters:
\al{
   n_s &= 0.96,\\
   r &= 0.060,\\
   n_r &= -0.00046,\\
   \Delta_R^2 &= 2.2\times 10^{-9	}\,\,\,. \label{eq:output1}
}
Fig.~\ref{fig:tilden21} displays the path followed by the inflaton during the 60 e-folds of inflation for this particular solution. The predictions in Eq.~(\ref{eq:output1})
are consistent with the results from the Planck experiment~\cite{Ade:2015lrj}:   $n_s=0.968 \pm 0.006$, $r<0.12$ (95\% C.L.), $n_r=-0.003 \pm 0.007$ and
$\Delta_R^2 = 2.19 \pm 0.08 \times 10^{-9}$. (The value of $\Delta_R^2$, also called $A_s$, was taken from the first column of Table~3 in Ref.~\cite{Ade:2015lrj}.)

We may check the validity of the results in this section by numerically evaluating the slow-roll parameters in the two-field problem.  Let $a$ represent the linear combination of
the fields that evolves along the minimum of the trench.   Given that $da =\sqrt{d\tilde{\rho}^2+d\tilde{\theta}^2}$ along the trench, it follows that we can write the $n^{th}$
derivative of the potential with respect to $a$ as
\begin{equation}
\frac{d^n V}{da^n} = \left[ \left(1+\frac{d\tilde{\rho}} {d\tilde{\theta}}\right)_{tr}^{-1/2} \frac{d}{d\tilde{\theta}} \right] ^n V\left(\tilde{\theta},\tilde{\rho}(\tilde{\theta})_{tr}\right) \,\,\, ,
\label{eq:numapprox}
\end{equation}
where the subscript ``tr" indicates quantities evaluated along $\tilde{\rho}(\tilde{\theta})_{tr}$, the solution to Eq.~(\ref{firstder1}).  Note that as the quantity $da$ is defined
above, the kinetic terms for $a$ are canonically normalized. The slow roll parameters can be evaluated numerically according to Eq.~(\ref{eq:numapprox}).  We find in this case 
that $n_s=0.96$, $r=0.060$, $n_r=-0.00046$ and $\Delta_R^2 = 2.2\times 10^{-9}$, in agreement with the results in Eq.~(\ref{eq:output1}).  

\subsection{Termination with a waterfall.} \label{n21-hybrid} 

For different choices of the model parameters, inflation will end before $\epsilon=1$ is reached, at a point where there is no longer a solution to Eq.~(\ref{firstder1}).  At this point,
the stability condition $\partial^2 V/\partial \tilde{\rho}^2 > 0$ is also not satisfied, and the fields evolve rapidly in a direction orthogonal to the original trajectory~\cite{Carone:2014cta}.  
If one visualizes the motion by plotting the fields as polar coordinates, the evolution corresponds to a transition from spiraling to rapid motion in the radial direction, eventually ending at a global 
minimum. In Ref.~\cite{Carone:2014cta} this was called the waterfall, in analogy to the behavior of hybrid inflation models~\cite{Linde:1993cn}, where stability in one field direction 
can be a function of the value of a second field.

\begin{figure}[ht]
\includegraphics[width = 0.5\textwidth]{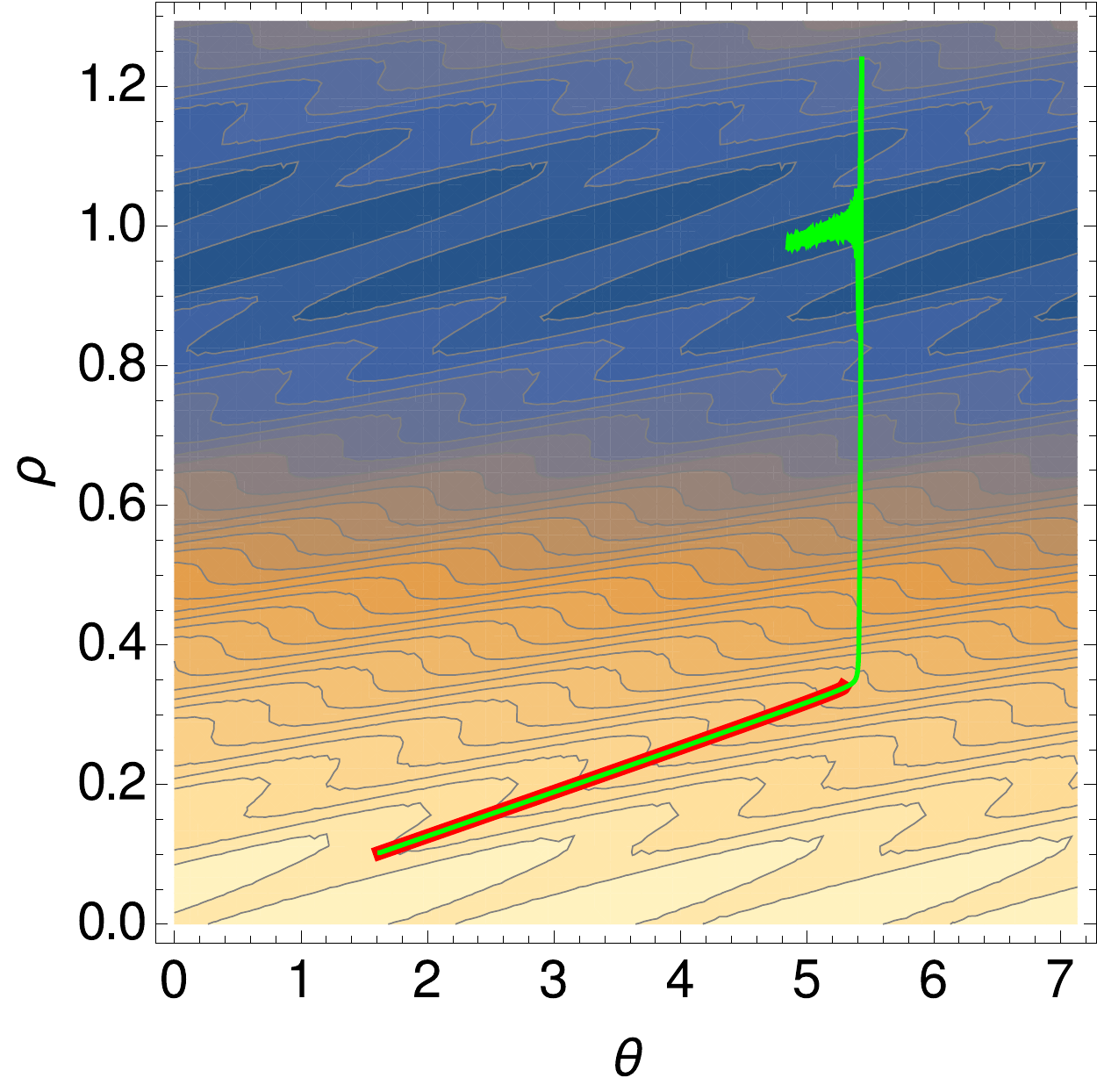}
\caption{Inflaton trajectory, in $\rho$-$\theta$ space, overlaid on a contour plot of the potential, in units where $M_*=1$. The bottom of the trench is indicated by the thick red line while the inflation trajectory is denoted 
by the thin green line.}
\label{fig:wf}
\end{figure}

Given an input of model parameters, we determine the final inflaton field value $a_f$  by solving
\begin{equation}
\frac{\partial^2V}{\partial \tilde{\rho}^2}\bigg|_{tr}=0 \,\,\, ,
\end{equation}
and then the initial value $a_i$ from
\begin{equation}
N=\int_{a_i}^{a_f} \!\bigg|\frac{V}{V'}\bigg| \, da \,\, .
\label{eq:ne}
\end{equation}
where the primes refer to derivatives evaluated numerically according to Eq.~(\ref{eq:numapprox}), and $a$ ($\approx \tilde{\theta}$) is the canonically normalized
inflaton field.  Again, we set $N=60$. To illustrate a solution that ends with the waterfall behavior, consider the parameter choices
\al{
  f_1 &= 0.22 \sqrt{2}\,\,\, , \\
  f_2 &= f_1 / 17 \,\,\, ,\\  
  \Lambda_1 & = 3.38 \times 10^{-3} \,\,\, , \\
  \Lambda_2 & = 1.61 \times 10^{-3} \,\,\, . 
}
which corresponds to $n=17$ and $\kappa =1.13$.  We find that the initial and final fields for the inflaton trajectory are given by
\begin{equation}
   (\rhotil,\thetil)_i =(6.83 \times 10^{-3},1.63) \,\,\,\,\,\,\,\, \mbox{ and } \,\,\,\,\,\,\,\, (\rhotil,\thetil)_f =(0.0281,  5.2970) \,\,\, ,
\end{equation}   
respectively.  Using this value for $\tilde{\theta}_i$, we find the following set of cosmological parameters:
\al{
   n_s &= 0.96,\\
   r &= 0.0078,\\
   n_r &= -7.2 \times 10^{-5},\\
   \Delta_R^2 &= 2.2\times 10^{-9	}\,\,\,. \label{eq:output2}
}
These are consistent with the ranges allowed by Planck, as quoted in the previous subsection. The complete inflaton trajectory, extending beyond the point 
where Eq.~(\ref{firstder1}) is no longer satisfied, can be found by solving the coupled equations of motion
\begin{eqnarray}
\ddot{\rho} + 3 H \dot{\rho} + \frac{\partial V}{\partial \rho}&=&0  \,\,\, ,\nonumber \\
\ddot{\theta} + 3 H \dot{\theta} + \frac{\partial V}{\partial \theta}&=&0 \,\,\, ,
\label{eq:eom}
\end{eqnarray}
where $H$ is the Hubble parameter. The result is shown in Fig.~\ref{fig:wf}. One can see from the plot that the bottom of the trench given by Eq.~(\ref{firstder1}), denoted by the thick red line, approximates the actual
trajectory, given by the thin green line, very well.  The inflaton eventually oscillates about and then settles at the global minimum of the potential.

\section{Models} \label{sec:fmodel}

\subsection{Origin of the potential} 

The successful inflation potentials presented in the previous section correspond to a potential of the form
\begin{equation}
V(\rho,\theta) = \Lambda_1^4 \left[1+\cos\left(\frac{\rho}{f_\rho}\right)\right] + \Lambda_2^4 \left[1-\cos\left(\frac{n \rho}{f_\rho} - \frac{\theta}{f_\theta}\right)\right] \,\,\, ,
\label{eq:pot2}
\end{equation}
where $n$ is an integer.   Here we consider the possibility that this potential arises via the effects of anomalies associated with new gauge groups.   

Hence, we extend the standard model gauge group by the additional factors SU($N_1$)$ \times$SU($N_2$), and introduce the fermions  
$A_L \sim A_R \sim ({\bf N_1}, {\bf 1})$ and $B_L^{(i)} \sim B_R^{(i)} \sim C_L \sim C_R \sim ({\bf 1}, {\bf N_2})$.  We would like the Lagrangian to contain the following interactions
\begin{equation}
{\cal L} \supset h_1 \bar{A}_R  A_L \chi  +\sum_{i=1}^n h_2^{(i)} \bar{B}_R^{(i)} B_L^{(i)} \chi + h_3 \bar{C}_R C_L \Phi^*+ \mbox{ h.c. } \,\,\, .
\label{eq:desired}
\end{equation}
Here, the $h_j$'s are Yukawa couplings and the terms shown generate heavy fermion masses when the $\Phi$ and $\chi$ fields develop vevs.  The accidental
global U(1) symmetries are each chiral when appropriate charges are assigned to the $A$, $B$ and $C$ fermions.  However, these symmetries are
anomalous with respect to the new gauge groups.  Triangle diagrams lead to the interactions~\cite{mcdonald}
\begin{equation}
\frac{g_1^2}{32 \pi^2} \left(\frac{\rho}{f_\rho}\right) F_1 \widetilde{F}_1 
+ \frac{g_2^2}{32 \pi^2} \left(\frac{n \rho}{f_\rho}-\frac{\theta}{f_\theta}\right) F_2 \widetilde{F}_2  \,\,\,,
\end{equation}
and the non-perturbative generation of a potential~\cite{mcdonald}
\begin{equation}
V(\rho,\theta) = \Lambda_1^4 \left[1-\cos\left(\frac{\rho}{f_\rho}\right)\right] + \Lambda_2^4 \left[1-\cos\left(\frac{n \rho}{f_\rho} - \frac{\theta}{f_\theta}\right)\right] \,\,\, ,
\end{equation}
with the scales $\Lambda_1$ and $\Lambda_2$ identified with the scale of strong dynamics for each SU($N$) factor. (We assume $N_1$ and $N_2$ are chosen so that each group is asymptotically 
free.)  Redefining the origin of field space via
\begin{equation}
\rho \rightarrow \rho + \pi f_\rho \,\,\,\,\,\,\,\,\,\, \mbox{ and } \,\,\,\,\,\,\,\,\,\, \theta \rightarrow \theta+n \pi f_\theta 
\end{equation}
puts the potential in the form that we previously assumed in Eq.~(\ref{eq:pot2}).  Note that the new gauge groups may be spontaneously broken at a scale well below
$\Lambda_1$ and $\Lambda_2$ without affecting our conclusions.

The interactions given in Eq.~(\ref{eq:desired}) are clearly not generic. In the absence of our discrete charge assignments for $\Phi$ and $\chi$, there would be
no reason for the $\Phi$ field to avoid coupling to the $A$ and $B$-type fermions directly, nor would there be any prohibition of explicit fermion mass terms.  Hence, this
sector is suggestive of additional symmetries even had we not put them forward immediately as a starting assumption in our model building.  Given the transformation
properties of $\Phi$ and $\chi$ fields under the $\mathbf{Z}^\Phi_p \times \mathbf{Z}^\chi_r$ symmetry, Eq.~(\ref{eq:rule}), we can account for the desired
pattern on couplings in Eq.~(\ref{eq:desired}) by choosing
\begin{equation}
A_R \rightarrow \omega_\chi A_R \,\,\, , \,\,\,\,\,\,\,\,\,\, B_R^{(i)} \rightarrow \omega_\chi B_R^{(i)} \,\,\, , \,\,\,\,\,\,\,\,\,\, C_L \rightarrow \omega_\Phi C_L \,\,\, ,
\end{equation}
with the remaining heavy fermions taken as singlets under the discrete group.  However, we must now enlarge the fermion content to assure that discrete gauge anomalies are cancelled (see the appendix), and do so in a way that assures that the additional fermions can become massive. To demonstrate that this can be 
accomplished, let us consider an example suggested by one of our previous cosmological solutions, discussed in Sec.~\ref{n21-nowaterfall}, corresponding
to the potential in Eq.~(\ref{eq:pot2}) with $n=21$.   Let us choose $p=r=21$.  First, we note that there are $21$ $B$-type fermions transforming 
each with $\mathbf{Z}_{21}^\chi$ charge $+1$, where we specify the charge $Q$ by defining the group element to be $\exp(2i \pi Q/21)$.  This implies that the
$\mathbf{Z}_{21}^\chi$-SU($N_2$)$^2$ discrete anomaly cancellation condition would be satisfied by the $B$ particle content alone.  The $A$ and $C$ fermions, on the
other hand, lead to anomalies,  so we include additional fermions with matching gauge quantum numbers and the discrete transformation rules
\begin{eqnarray}
&&A^{(i)}_R \rightarrow \omega_\chi^{10} A^{(i)}_R  \,\,\,\,\,\,\,\,\,\, , \,\,\,\,\,\,\,\,\,\, A^{(i)}_L \rightarrow A^{(i)}_L \,\,\,\,\,\,\,\,\,\, (i=1\dots 2)  \nonumber \\
&&C^{(i)}_L \rightarrow \omega_\Phi^{10} C^{(i)}_L  \,\,\,\,\,\,\,\,\,\, , \,\,\,\,\,\,\,\,\,\, C^{(i)}_R \rightarrow C^{(i)}_R \,\,\,\,\,\,\,\,\,\, (i=1\dots 2) 
\end{eqnarray}
which allow the anomaly cancellation conditions to be satisfied.   Finally, we note that these fields will develop masses as a result of Planck-suppressed operators, for example,  $\bar{A}^{(i)}_R \chi^{10}  A^{(i)}_L / M_*^9 + \mbox{ h.c.}$ and $\bar{C}^{(i)}_L \Phi^{10}  C^{(i)}_R / M_*^9 + \mbox{ h.c. }$,
which lead to masses of order $\lambda^{10} M_* \sim 10^{11}$~GeV.   

The discrete symmetry that we have assumed to assure the form of couplings in Eq.~(\ref{eq:desired}) also leads to a suppression of direct Planck suppressed corrections 
to the potential.  Since quantum gravitational effects must respect the discrete gauge symmetry, the lowest order operators that will correct the potential have the form $
\Phi^{21}/M_*^{17}$ or $\chi^{21}/M_*^{17}$; the scale of these corrections are of order $ \lambda^{21} M_*^4 \sim 10^{-14} M_*^4$,  negligible compared to the 
values of $\Lambda_1$ and $\Lambda_2$ that we found previously to be of order $10^{-3} M_*$.

\subsection{Standard Model Flavor}

The fields $\Phi$ and $\chi$ can now be utilized in constructing models of standard model fermion masses.  These fields will
appear in higher-dimension operators that generate the small entries of the standard model Yukawa matrices.  Given our choice
$\langle \Phi \rangle / M_* = \langle \chi \rangle/M_*=\lambda$, the size of these entries will be determined by powers of the
Cabibbo angle $\lambda$.  In this subsection, we present one example in which the desired set of higher-dimension operators is
obtained via the same discrete symmetries that were used to obtain the inflaton potential.  We focus on the $n=p=r=21$ 
model just discussed, in which the $\Phi$ and $\chi$ fields each transform under a separate $\mathbf{Z}_{21}$ symmetry.  
Of course, other choices of the symmetry group are possible, and the present choice does not suggest a unique set of
fermion charge assignments (since there are many possible Yukawa textures that are viable).   The example we give here 
will suffice by serving as a proof of principle\footnote{It should also be clear that one could alternatively construct a model starting 
with the $n=17$ potential that we discussed earlier, but there are no new qualitative insights gained by presenting two very similar examples.}.

The simplest incorporation of the $n=21$ model in a flavor sector is via the identification of $\mathbf{Z}^\Phi_{21}$ as the flavor symmetry and
$\Phi$ as the sole flavon field.   The charge assignments of the standard model fermions and a set of right-handed neutrinos are given in Table~\ref{charges}.
\begin{table}[t]
\begin{tabular}{c ccc c ccc c ccc c ccc c} \hline\hline
$Q_{1L}$ && $Q_{2L}$ && $Q_{3L}$ && $u_R^c$ && $c_R^c$ && $t_R^c$ && $d_R^c$ && $s_R^c$ && $b_R^c$  \\
\hline
6 && 5 && 3 && 2 && -1 && -3 && -1 && -2 && -2 \\ \hline \\ \hline\hline
$L_{1L}$ & & $L_{2L}$ & & $L_{3L}$ & & $e_R^c$ & & $\mu_R^c$ & & $\tau_R^c$ && $\nu_{1R}^c$ && $\nu_{2R}^c$ && $\nu_{3R}^c$ \\
\hline
$0$ && $0$ && $0$ && $5$ && $3$ && $1$ && $-3$ && $-3$ && $-3$ \\ \hline\hline
\end{tabular}
\caption{$\mathbf{Z}^\Phi_{21}$ charge assignments $q$, where the group transformation is defined by $\exp(2 i \pi q/21)$. The Higgs doublet is a singlet under 
the flavor symmetry.} \label{charges}
\end{table}
Entries of the Yukawa matrices arise from $\mathbf{Z}^\Phi_{21}$-invariant higher dimension operators.  For example, the $1$-$1$ entry in the up-sector 
Yukawa matrix involves the fields $\overline{Q}_{1L} H u_R$, which has flavor charge $-8$.  This arises at lowest order via
\begin{equation}
\frac{1}{M_*^8}\overline{Q}_{1L} H \Phi^8 u_R + \mbox{ h.c.} \,\, ,
\end{equation}
and hence the corresponding Yukawa matrix entry is of order $\lambda^8$.  Since $\omega^8$ and $\omega^{-13}$ are identical, there is another possible
operator, $\overline{Q}_{1L} H \Phi^{* 13} u_R/M_*^{13} + \mbox{ h.c.}$, but it is of higher order and can be neglected.  We may populate the remaining entries of the quark 
and charged lepton Yukawa matrices in a similar manner.  We find
\begin{equation}
Y_u  = \left(\begin{array}{ccc} \lambda^8 & \lambda^5 & \lambda^3 \\
\lambda^7 & \lambda^4 & \lambda^2 \\
\lambda^5 & \lambda^2 & 1 \end{array}\right)
\,\,\,\,\,\,\,\, , \,\,\,\,\,\,\,\,
Y_d  = \left(\begin{array}{ccc} \lambda^5 & \lambda^4 & \lambda^4 \\
\lambda^4 & \lambda^3 & \lambda^3 \\
\lambda^2 & \lambda & \lambda \end{array}\right)
\,\,\,\,\,\,\,\, , \,\,\,\,\,\,\,\,
Y_e  = \left(\begin{array}{ccc} \lambda^5 & \lambda^3 & \lambda \\
\lambda^5 & \lambda^3 & \lambda \\
\lambda^5 & \lambda^3 & \lambda \end{array}\right)  \,\,\, ,
\end{equation}
where order one coefficients in each entry have been suppressed.  These achieve the desired ratios $m_u/ m_t \sim \lambda^8$, $m_c/m_t \sim \lambda^4$,
$m_d/m_b \sim \lambda^4$, and $m_s/m_b \sim \lambda^2$, with the charged lepton Yukawa mass eigenvalues comparable in size to those of the 
down quark sector.  It is not hard to verify that the choice of right-handed neutrino charge assignments leads via the see-saw mechanism to a neutrino 
mass matrix of the form  $[\langle H \rangle^2/\Lambda_R]Y_\nu$, where $\Lambda_R$ is the right-handed neutrino mass scale, $\langle H \rangle$ is the
standard model Higgs vev, and $Y_\nu$ is a matrix in which each entry is of order $\lambda^0$ times a function of (typically many) undetermined 
order one coefficients. These can be chosen to obtain the desired phenomenology without unnaturally large or small values of the individual coefficients\footnote{It is 
not necessarily the case that an alternative model that predicts the neutrino mass hierarchy via powers of $\lambda$ is more desirable than this example.  The reason 
is that the predictions for neutrino mass matrix entries in such a model also come multiplied by functions of products of a number of the order one operator coefficients.
This can spoil the naive $\lambda$ power counting without any individual operator coefficient being unnaturally small or large.  This is a problem that is unique to the 
neutrino sector in such models when the mass matrix arises via the seesaw mechanism.}.

Finally, we must check that the standard model fermion charge assignments in this model satisfy the linear Ib\'{a}\~{n}ez-Ross anomaly cancellation conditions
for the non-Abelian gauge groups and gravity. Summing the $\mathbf{Z}^\Phi_{21}$ charges times the appropriate multiplicity factors for the color SU(3), weak SU(2), and 
gravitational anomalies gives $21$, $42$ and $63$, respectively.  These results mod $21$ are zero, indicating that the discrete gauge anomaly cancellation conditions discussed
in Appendix~\ref{sec:appendix} remain satisfied.

\section{Conclusions} \label{sec:conc}

Models of standard model flavor that are based on discrete gauge symmetries can have accidental continuous global symmetries that are
spontaneously broken.  We have argued that a linear combination of the approximate goldstone bosons that may arise in these models can serve 
plausibly as the inflaton in two-field models of inflation based on the axion monodromy idea.  These models can accommodate the current Planck data
on the microwave background~\cite{Ade:2015lrj} while allowing the flavor-symmetry-breaking vacuum expectation values (vevs) to remain sub-Planckian.  This is
important in the present work since the ratios of the flavon vevs to the reduced Planck scale serve as small flavor-symmetry-breaking parameters in our models,
which allows one to predict the standard model Yukawa coupling entries in a controlled approximation.   In addition to making correct Yukawa coupling 
predictions possible, the discrete symmetries of the theory also maintain the correct pattern of the interactions in a new gauge sector, leading to 
the desired form of the inflaton potential; they also keep the quantum gravitational corrections to the potential well under control.  The 
literature on models of standard model fermion masses is vast and it is imaginable that more economical and  compelling examples of flavor-sector inflation 
models are yet to be found.  The present work suggests that exploring the full landscape of such models may be a fertile direction for future investigation.

\begin{acknowledgments}  
This work was supported by the NSF under Grants PHY-1068008 and PHY-1519644.  The authors thank Joshua Erlich for valuable
discussions.
\end{acknowledgments}

\appendix
\section{Discrete Gauge Symmetries, Briefly} \label{sec:appendix}
It is well known that continuous gauge symmetries are not violated by quantum gravitational effects. Under what circumstances is the
same true for discrete symmetries?  It was noted long ago by Ib\'{a}\~{n}ez and Ross (IR)~\cite{Ibanez:1991hv} that a discrete group that arises as a subgroup of
a continuous gauge symmetry inherits this protection.  While the full theory must satisfy the anomaly cancellation conditions relevant
for the continuous gauge groups, IR determined the conditions that are relevant in the low-energy theory, below the scale at
which the continuous gauge symmetries are broken.  Since some of the fermions in the complete theory may become massive and decoupled when symmetry
breaking occurs, the low-energy theory includes only part of the fermion content that contributes to anomaly cancellation in the full theory.  The low-energy
constraints should refer only to the light fermion content, which in the present context corresponds to models defined below the reduced Planck scale $M_*$.
If the appropriate consistency conditions are satisfied, the discrete gauge symmetry can be treated as fundamental, without reference to specific high-energy
embeddings.

The constraints that we apply in our model building are the linear IR conditions involving non-Abelian gauge group factors; these follow from triangle diagrams involving two non-Abelian gauge group factors and one factor of the continuous gauge group in which the discrete symmetry is embedded.  For example, the $\mathbf{Z}_N$-SU($M$)$^2$ anomaly cancellation condition is~\cite{Ibanez:1991hv}
\begin{equation} 
\sum_i C_i \, q_i = \frac{1}{2}\, r\, N \,\,\, .
\end{equation}
Here $r$ is an integer, $q_i$ is the $\mathbf{Z}_N$ charge of the $i^{th}$ fermion (which transforms under $\mathbf{Z}_N$ by $\exp[i 2\pi q_i/N]$) and $C_i$
is the Casimir invariant given by $\mbox{Tr} (T^a T^b)=C_i \delta^{ab}$, where the $T^a$ are SU($M$) generators in the representation of the $i^{th}$ fermion.  Since 
all the fermions in the model presented in Sec.~\ref{sec:fmodel} are in the fundamental representations of the relevant SU($M$) gauge groups, $C_i=1/2$; the 
linear IR conditions simply requires that the $\mathbf{Z}_N$ charges of the fermions that transform under a specified SU($M$) factor sum to an integer multiple of $N$.   
According to IR, when $N$ is odd (relevant to the model of Sec.~\ref{sec:fmodel}) the gravitational anomalies linear in $\mathbf{Z}_N$ are cancelled when the sum of all 
the $\mathbf{Z}_N$ charges are also an integer multiple of $N$.  It is straightforward to verify that these conditions are satisfied by the charge assignments displayed in
Table~\ref{charges}.

What about the other possible anomaly cancellation conditions?   First, IR note that the linear conditions involving the Abelian gauge groups do not lead to any useful constraints
on the low-energy theory~\cite{Ibanez:1991hv}.  Banks and Dine (BD)~\cite{Banks:1991xj} later showed that the IR conditions non-linear in the discrete group
make a tacit assumption about the high-energy embedding of the theory, through the requirement that both the light and the heavy fermions have 
integer U(1) charges.   BD show that there are consistent,  non-anomalous theories  (ones in which the effective discrete symmetry at low energies is smaller than that of the 
full theory) in which the low-energy spectrum does not satisfy the non-linear IR constraints; their failure only implies the existence of heavy fermions with fractional
charges.  Thus, the non-linear IR conditions are not required for the consistency of the low-energy effective theory.   BD note that the surviving discrete anomaly cancellation conditions are
physically sensible:  for example, the condition for the cancellation of the  $\mathbf{Z}_N$-SU($M$)$^2$ anomaly also guarantees that there are no t'Hooft interactions generated 
by SU($M$) instantons that would explicitly break  the $\mathbf{Z}_N$ symmetry.  This physical constraint~\cite{Preskill:1991kd} is completely independent of the high-energy embedding.

\end{document}